\documentclass[letterpaper]{article}

\usepackage{emulateapj}
\usepackage{onecolfloat}
\usepackage{apjfonts}
\usepackage{amsmath}
\usepackage{natbib}
\usepackage{graphicx}

\newcommand{\Msun}      {\mbox{$\rm\,M_{\mathord\odot}$}}
\newcommand{\Rsun}      {\mbox{$\rm\,R_{\mathord\odot}$}}

\begin{document}

\submitted{Accepted by ApJ Letters}

\twocolumn[
\title{Keck Measurement of the XTE~J2123--058 Radial Velocity Curve}

\author{John A. Tomsick and William A. Heindl}
\affil{Center for Astrophysics and Space 
Sciences, University of California, San Diego, MS 0424, La Jolla, CA 92093}

\author{Deepto Chakrabarty}
\affil{Department of Physics and Center for Space Research, 
Massachusetts Institude of Technology, Cambridge, MA 02139}

\author{Jules P. Halpern}
\affil{Columbia Astrophysics Laboratory, Columbia University, 
550 West 120th Street, New York, NY 10027}

\author{Philip Kaaret}
\affil{Harvard-Smithsonian Center for Astrophysics, 60 Garden Street, 
Cambridge, MA 02138}

\begin{abstract}

We measured the radial velocity curve of the companion of the neutron
star X-ray transient XTE~J2123--058.  Its semi-amplitude ($K_{2}$) of 
$298.5\pm 6.9$~km~s$^{-1}$ is the highest value that has been measured 
for any neutron star LMXB.  The high value for $K_{2}$ is, in part, due 
to the high binary inclination of the system but may also indicate a 
high neutron star mass.  The mass function ($f_{2}$) of $0.684\pm 0.047$\Msun,
along with our constraints on the companion's spectral type (K5V-K9V)
and previous constraints on the inclination, gives a likely range
of neutron star masses from 1.2 to 1.8\Msun.  We also derive a source 
distance of $8.5\pm 2.5$~kpc, indicating that XTE~J2123--058 is unusually 
far, $5.0\pm 1.5$~kpc, from the Galactic plane.  Our measurement of the 
systemic radial velocity is $-94.5\pm 5.5$~km~s$^{-1}$, which is significantly 
different from what would be observed if this object corotates with the disk 
of the Galaxy.

\end{abstract}


] % twocolumn

\section{Introduction}

The X-ray transient XTE~J2123--058 is a neutron star low mass X-ray binary 
(LMXB) that was discovered in 1998 \citep{lss98}.  We optically 
identified the transient \citep{tomsick98a} and determined the $\sim 6$~hour 
binary orbital period \citep{tomsick98b}.  Type I X-ray bursts indicate that 
the system contains a neutron star, and the relatively short orbital period 
indicates that the companion is a late-type star on or close to the main 
sequence.  A pair of high frequency quasi-periodic oscillations was detected, 
which may indicate a rapidly rotating neutron star \citep{homan99,tomsick99}.  
XTE~J2123--058 distinguishes itself from other LMXBs by having high Galactic 
latitude ($b = -36^{\circ}$) and also a high binary inclination 
($i\sim 73^{\circ}$, Zurita et al.~2000\nocite{zurita00}).  Both of these 
properties are advantageous for determining the mass of the neutron star from 
optical observations in quiescence, which is the primary goal of this work.

Such mass measurements are important for constraining neutron star equations 
of state and for understanding the evolution of neutron star systems.  
Precise mass measurements have been made for millisecond 
radio pulsars (MSPs, Thorsett \& Chakrabarty 1999\nocite{tc99}), but these 
measurements are lacking for LMXBs.  Since it is theoretically possible to 
spin-up neutron stars in LMXBs to millisecond periods via accretion, 
a link between LMXBs and MSPs has long been suspected \citep{alpar82}.  
Although there is substantial evidence to support this picture, the prediction 
that rapidly rotating neutron stars in LMXBs should be more massive 
by 0.1 to 0.5\Msun~(Bhattacharya 1995\nocite{bhattacharya95}) than those 
that have not been spun-up has not been tested.

\section{Observations and Data Reduction}

In 2000 August, we performed spectroscopy of XTE~J2123--058 with the Echelle 
Spectrograph and Imager (ESI, Sheinis et al. 2000\nocite{sheinis00}) at Keck 
Observatory.  As shown in Table~\ref{tab:obs}, we took eight exposures over 
two nights covering the 6~hour binary orbit.  The cross-dispersed spectra were 
taken in Echelle mode with ten spectral orders falling on a 2048-by-4096 pixel 
CCD.  We performed on-chip rebinning to 2048-by-2048 pixels to decrease read 
noise.  Although this instrument is sensitive from 3900-11000\AA, we only used 
the data in the 4700-6820\AA~band where the dispersion varies 
from 0.36 to 0.52\AA~per rebinned pixel.  This is sufficient to over-sample the 
resolution of the spectrograph as we used a 1$^{\prime\prime}$ slit, providing 
a spectral resolution of 1.0 to 1.5\AA~FWHM.

We obtained R and V-band photometry of XTE~J2123--058 at MDM Observatory 10~days 
before the Keck observations, and found values for R and V consistent with the 
quiescent magnitudes reported in \cite{tomsick99}.  We measured 
$V = 22.68\pm 0.11$ (corrected for atmospheric extinction but not reddening), 
and we use this value below to determine the source distance.  Due to the 
faintness of the source, relatively long exposure times between 2340 and 2700 
seconds were necessary for our spectroscopic observations.  Even these exposures 
are a relatively small fraction of the orbital period.  We obtained M, K and G 
dwarf comparison star spectra for the cross-correlations described below.  

We extracted the spectra using the software package ``makee''
which was originally developed for the HIRES instrument at Keck, 
but has been adapted to extract ESI spectra.  Using makee, we
produced optimally extracted spectra for each of the ten spectral 
orders.  We obtained CuAr line lamp exposures to determine the 
wavelength solution, and we checked this solution for each exposure 
using OI and NaI night sky lines.  Averaging over all exposures and 
night sky lines, the difference between measured and actual 
wavelengths is 0.018\AA.  The calibration is also stable from 
exposure-to-exposure with a standard deviation of 0.027\AA~or 
less for all night sky lines, corresponding to a radial velocity 
measurement error of less than 1.5~km~s$^{-1}$.  The data reduction 
procedure includes heliocentric velocity and atmospheric extinction 
corrections.  We used IRAF routines to flux calibrate the spectra 
and obtained exposures of the spectrophotometric standard 
Hiltner~102 \citep{stone77} on both nights for this purpose.  We 
combined the spectral orders after flux calibration and rebinned 
the spectra to a logarithmic wavelength scale with 22.8~km~s$^{-1}$ 
pixels.  Finally, we dereddened the XTE~J2123--058 spectra using 
$A_{V} = 0.37$~magnitudes \citep{hynes01}.

\begin{table}[t]
\caption{XTE~J2123--058 Keck Observations\label{tab:obs}}
\begin{minipage}{\linewidth}
\footnotesize
\begin{tabular}{c|c|c|c} \hline \hline
 & MJD\footnote{Modified Julian Date (JD$-$2400000.5) at the midpoint of the exposure.}
& Exposure Time & FWHM Seeing\\
Exposure & (days) & (seconds) & (arcseconds)\\ \hline
1 & 51759.39442 & 2700 & 0.87\\
2 & 51759.42822 & 2510 & 0.91\\
3 & 51759.45771 & 2510 & 1.23\\
4 & 51759.48768 & 2510 & 0.79\\
5 & 51759.51748 & 2400 & 0.85\\
6 & 51759.54536 & 2340 & 0.98\\
7 & 51759.57288 & 2340 & 0.86\\
8 & 51760.35770 & 2400 & 0.67\\
\end{tabular}
\end{minipage}
\end{table}

The quality of the XTE~J2123--058 spectra varies significantly
between exposures as indicated by the signal-to-noise ratios (SNR) 
given in Table~\ref{tab:rv}.  The mean SNR per pixel in the
5650-5850\AA~band is 1.66 in the worst case and 6.63 
in the best case.  One reason for this is that the source 
shows large orbital flux variations as can be seen from the fluxes
given in Table~\ref{tab:rv}.  The flux variations show two peaks
per orbit (exposures 5 and 8), which is expected if the variability
is due to ellipsoidal modulations.  XTE~J2123--058 light 
curves from optical photometry show that ellipsoidal modulations
are indeed present \citep{zurita00}.  The SNR variations are also 
caused by variable seeing as shown in Table~\ref{tab:obs}.  The 
FWHM seeing for the exposure taken on the second night is 
$0^{\prime\prime}.67$ compared to an average of 
$0^{\prime\prime}.93$ for the exposures on the first night.  It 
should be noted that the Table~\ref{tab:rv} fluxes are not 
corrected for seeing.

\section{Results}

\subsection{Cross-Correlations}

Our analysis of the spectra focuses on the region from 4700\AA~to 
6820\AA~since the SNR becomes very poor below 4700\AA~and night sky 
lines dominate the spectrum above 6820\AA.  In this band, the spectra 
show features from the accretion disk (H$\alpha$ and H$\beta$ emission 
lines) and from the companion star.
An examination of the spectra clearly shows the presence of a wide 
depression centered between 5100 and 5200\AA.  This is the signature of 
a K-type star and is due to the MgH feature at 5180\AA, the TiO band at 
4954-5200\AA~and a continuum discontinuity associated with the MgI 
triplet at 5168-5185\AA~(McClintock \& Remillard 1990\nocite{mr90} and 
references therein).  An absorption feature is seen near 5900\AA~due to 
the NaI doublet at 5891-5898\AA.  The strength of this feature 
indicates a late-K spectral type for the companion.  

\begin{figure}[t]
\plotone{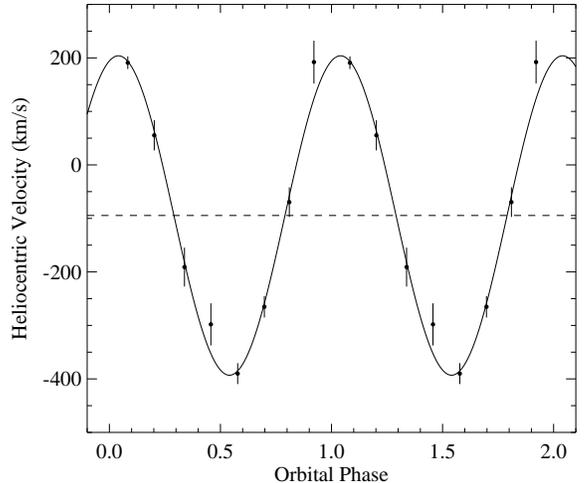}
\caption{Radial velocity curve fitted using a sinusoid with the 
parameters given in Table~\ref{tab:parameters} and folded on the 
orbital period.  Heliocentric velocities are shown (with errors 
increased by 60\% over the values given in Table~\ref{tab:rv}), 
and the dashed line marks the systemic velocity.  The data
are plotted twice for clarity.
\label{fig:rv}}
\end{figure}

We used the IRAF routine ``fxcor'' \citep{td79,fmh95} to cross-correlate 
the XTE~J2123--058 spectra with the comparison stars.  After masking
night sky lines at 5205, 5579, 5891, 5898 and 6302\AA~and the
H$\alpha$ and H$\beta$ emission lines from the source, we used
the wavelength bands 4890-5190, 5210-5560, 5630-5860, 5985-6135 and 
6640-6820\AA~for the cross-correlations.  We also applied a bandpass 
filter from 30 to 800 cycles per spectrum in order to remove contributions 
due to spurious variations in the continuum and high frequency noise.
We calculated cross-correlations for G0V, G5V, G9V, K0V, K2V, K4V,
K7V, M0V and M1V with the exposure 8 spectrum.  Highly significant 
peaks with Tonry \& Davis (1979)\nocite{td79} R-values between 8.9 and 16.3 
at roughly the same velocity occur, with the M-type and G0V stars giving the 
weakest correlations and the K7V star (HD~88230) giving the strongest 
correlation.  For this reason and because of the strong NaI line, we used 
the K7V star for the radial velocity study.

\begin{table}[b]
\caption{Fluxes, Signal-to-Noise Ratios and Cross-Correlation Results\label{tab:rv}}
\begin{minipage}{\linewidth}
\footnotesize
\begin{tabular}{c|c|c|c|c|c} \hline \hline
Exposure & Flux\footnote{Mean flux in the 5650 to 5850\AA~band in units of 
$10^{-18}$~erg~cm$^{-2}$~s$^{-1}$~\AA$^{-1}$.} & SNR\footnote{Mean signal-to-noise
ratio per pixel in the 5650 to 5850\AA~band.} & R\footnote{\cite{td79} R-value} & 
Measured RV\footnote{Measured heliocentric radial velocity in km~s$^{-1}$ with
\cite{td79} errors.} & Fitted RV\footnote{Heliocentric radial velocity in km~s$^{-1}$ 
based on a sinusoidal fit to the data.}\\ \hline
1 & 1.24 & 1.66 & 3.26 & $55.3\pm 17.6$ & $63.2$\\
2 & 1.72 & 2.66 & 4.29 & $-191.0\pm 22.7$ & $-182.4$\\
3 & 1.53 & 2.15 & 4.54 & $-298.1\pm 24.6$ & $-352.8$\\
4 & 2.90 & 4.40 & 8.88 & $-390.1\pm 12.2$ & $-385.0$\\
5 & 3.06 & 4.18 & 8.82 & $-265.2\pm 12.5$ & $-259.0$\\
6 & 2.48 & 2.87 & 5.14 & $-69.6\pm 17.3$ & $-58.1$\\
7 & 2.32 & 2.41 & 3.35 & $192.2\pm 25.0$ & $123.6$\\
8 & 5.53 & 6.63 & 16.32 & $191.0\pm 7.4$ & $193.5$\\
\end{tabular}
\end{minipage}
\end{table}

\begin{figure*}[t]
\centerline{\includegraphics[]{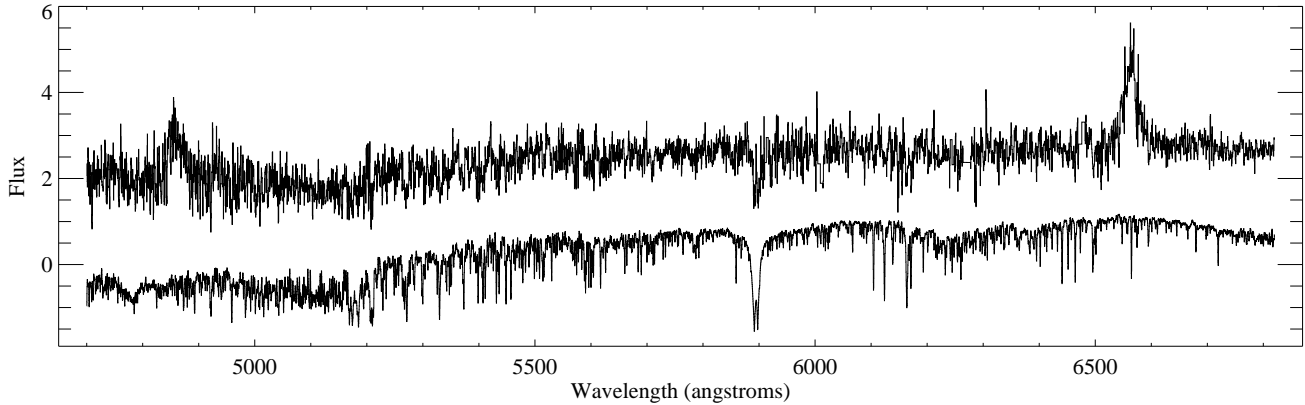}}
\caption{The top spectrum shows the average, Doppler corrected 
spectrum for XTE~J2123--058 in units of 
$10^{-18}$~erg~cm$^{-2}$~s$^{-1}$~\AA$^{-1}$.  H$\alpha$ and
H$\beta$ emission lines are present at 6564\AA~and 4863\AA.
The spectrum for the K7V comparison star (HD~88230) is shown below 
and the units are arbitrary.  Several matching features between
XTE~J2123--058 and HD~88230 are present including the continuum
break near 5200\AA~and the NaI doublet at 5891-5898\AA.
\label{fig:spectrum}}
\end{figure*}

We obtained unique and significant cross-correlation peaks at reasonable 
velocities ($-1000$ to 1000~km~s$^{-1}$) for all eight exposures.  The 
measured heliocentric velocities range from $-390$~km~s$^{-1}$ to 
192~km~s$^{-1}$ and are given in Table~\ref{tab:rv}.  The R-values vary 
significantly from 3.3 to 16.3 due to SNR differences.  Figure~\ref{fig:rv} 
shows the radial velocity curve.  First, we performed a least-squares fit to 
the data with a function consisting of a constant plus a sinusoid with four 
free parameters:  the systemic velocity ($\gamma$), the epoch of inferior
conjunction ($T_{0}$), the radial velocity semi-amplitude ($K_{2}$) and 
the orbital period ($P_{orb}$).  This does not provide a formally acceptable 
fit ($\chi^{2}/\nu = 12/4$) so we increased the errors on the individual
velocity measurements by 60\% to give a reduced-$\chi^{2}$ of 1.0.
The fit gives a large range of possible values for $P_{orb}$ with a 90\% 
confidence error region from 21125 to 22575~seconds.  This range contains 
the photometric orbital period of $21445.5\pm 2.3$~seconds measured 
previously \citep{tomsick99,ic98,zurita00}, and we refitted the radial velocity 
curve after fixing $P_{orb}$ to this value.  We expect that the photometric 
and spectroscopic orbital periods are the same since the photometric value is 
based on partial eclipses seen in the optical light curve during 
outburst \citep{tomsick99,zurita00}.  

Table~\ref{tab:parameters} shows the fit parameters with $P_{orb}$ fixed, 
and this fit is shown in Figure~\ref{fig:rv}.  Its semi-amplitude
($K_{2}$) of $298.5\pm 6.9$~km~s$^{-1}$ is the highest value
that has been measured for any neutron star LMXB.  The value of 
$-94.5\pm 5.5$~km~s$^{-1}$ for $\gamma$ is consistent with the tentative 
value of $-90$~km~s$^{-1}$ derived from emission lines \citep{hynes01}.  
We note that previous measurements of the photometric epoch are too 
uncertain to make a meaningful comparison to our spectroscopic epoch.

\subsection{Average Spectrum and Source Distance}

We produced an average, Doppler corrected spectrum using the eight 
XTE~J2123--058 exposures.  We first removed pixels contaminated by night 
sky lines.  Next, we multiplied each spectrum by a constant based on the 
fluxes given in Table~\ref{tab:rv} to establish a common normalization.  
This is important because the wavelength bins near night sky lines do not 
have contributions from all eight exposures.  We then Doppler corrected 
the spectra using the fitted radial velocity values given in 
Table~\ref{tab:rv}.  Finally, we interpolated the spectra to a common 
set of wavelength bins and calculated the weighted average for each 
wavelength bin.  

The average spectrum is shown in Figure~\ref{fig:spectrum} along
with the spectrum of HD~88230.  Gaussian fits to the XTE~J2123--058 
emission lines give equivalent widths (EW) of 15.8 and 17.6\AA~and 
FWHM widths of $1443\pm 64$ and $1729\pm 133$~km~s$^{-1}$ for 
H$\alpha$ and H$\beta$, respectively.  The FWHM widths are significantly 
less for the individual exposures.  The continuum features near 
5000-5300\AA~are similar to HD~88230, and narrow MgI absorption lines are 
seen in both spectra near the break in the continuum.  Matching absorption 
lines are also present in several other parts of the spectrum, including 
the NaI doublet at 5891-5898\AA.  

We fitted the region of the spectrum near the NaI doublet with a constant plus 
two inverted Gaussians for the XTE~J2123--058 average spectrum and also for 
the comparison stars in order to determine the total EW of the NaI doublet.  
For XTE~J2123--058, we measure an EW of $3.1\pm 0.6$\AA~(after correcting 
for interstellar absorption using results from Hynes et al.~2001\nocite{hynes01}).  
For the K0V, K2V, K4V, K7V and M1V comparison stars, we measure EW
of 1.1, 2.5, 3.6, 6.9 and 7.2\AA, respectively.  Since the H$\alpha$ and H$\beta$ 
emission lines indicate a significant disk contribution, these EW values indicate 
that the spectral type of the companion is later than K4V.  However, based on 
the strengths of the cross-correlation peaks described above, the companion is a 
K rather than an M-type star.  The range of possible spectral types is K5V-K9V.  

\begin{table}[b]
\caption{Spectroscopic Parameters\label{tab:parameters}}
\begin{minipage}{\linewidth}
\footnotesize
\begin{tabular}{c|c} \hline \hline
Parameter & Result\footnote{68\% confidence errors are given.}\\ \hline
Epoch\footnote{Heliocentric Julian Date of the epoch of inferior conjunction.} 
& $T_{0} =$ HJD $2451760.0462\pm 0.0013$~days\\
Systemic Velocity & $\gamma = -94.5\pm 5.5$~km~s$^{-1}$\\
Velocity Semi-Amplitude & $K_{2} = 298.5\pm 6.9$~km~s$^{-1}$\\
Orbital Period\footnote{Orbital period determined from photometry \citep{tomsick99,ic98,zurita00}.}
& $P_{orb} = 21445.5$~seconds\\
\end{tabular}
\end{minipage}
\end{table}

We determined the source distance using the method described by \cite{mcclintock01}.  
We assume a spectral type of K7V in the calculation and discuss possible 
associated uncertainties below.  The strength of the NaI doublet indicates that
the fraction of the light coming from the companion at 5900\AA~is 45\%$\pm$ 8\%.  
This, along with the MDM measurement (V = 22.68), gives an effective V-magnitude 
for the companion of 23.56.  We calculated the absolute V-magnitude ($M_{V}$) of 
the companion using the expression given in \cite{popper80}, which depends on the 
spectral type and the radius of the companion.  Since the companion fills its Roche 
lobe, its average density depends only on the orbital period \citep{ffw72}.  
From the density and assuming that the companion mass is 0.4\Msun, which is 
likely correct to within a factor of two \citep{bmg96}, we estimate a radius of 
0.56\Rsun~for the companion.  This radius and the K7V spectral type gives 
$M_{V} = 8.42$.  After accounting for extinction, we obtained a source distance 
of 9.0~kpc.  Recalculating with different spectral types indicates that the error 
associated with the K5V-K9V range is $\pm 1$~kpc.  Also, the factor of two 
uncertainty in the companion's mass translates to a 25\% uncertainty in the 
distance \citep{mcclintock01}.  Finally, there is a systematic error due to 
orbital variations in the V-magnitude.  The MDM observations were made close 
to photometric minimum so that the average V-magnitude is somewhat less than 22.68.  
Based on the quiescent light curve shown in \cite{zurita00}, we estimate that the 
average V-magnitude is close to 22.55, which decreases the distance by 0.5~kpc.
We conclude that the distance is $8.5\pm 2.5$~kpc, indicating that 
XTE~J2123--058 is $5.0\pm 1.5$~kpc from the Galactic plane.

\section{Discussion and Conclusions}

\begin{figure}[t]
\plotone{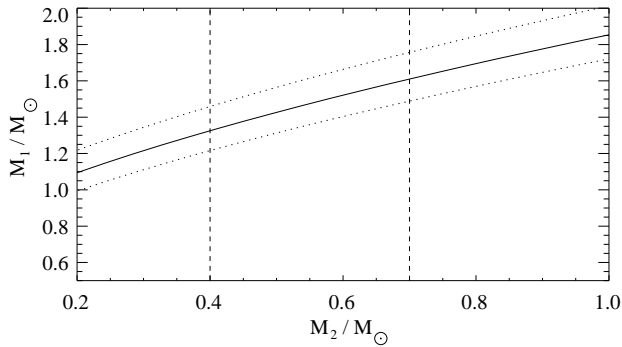}
\caption{The neutron star mass ($M_{1}$) vs.~the companion
mass ($M_{2}$).  The solid line comes from the values of 
$f_{2}$ and $i$ given in the text, and the dotted lines
are based on the extremes of the error regions for these
parameters.  The vertical dashed lines mark the likely
range for $M_{2}$.\label{fig:m1m2}}
\end{figure}

From $K_{2}$ and $P_{orb}$, we derive a mass function 
of $f_{2} = 0.684\pm 0.047$\Msun, representing a lower limit
on the neutron star mass ($M_{1}$).  The binary inclination 
($i$) and the mass ratio ($q = M_{1}/M_{2}$) are also necessary
to obtain a precise measurement of $M_{1}$.  For XTE~J2123--058, 
$i = 73^{\circ}\pm 4^{\circ}$ was determined via modeling
of the outburst optical light curves \citep{zurita00}.  
Figure~\ref{fig:m1m2} shows $M_{1}$ vs.~$M_{2}$ using the 
measured values of $f_{2}$ and $i$.  From our constraint on 
the spectral type of the companion (K5V-K9V), it is very likely 
that $M_{2}$ lies between 0.4 and 0.7\Msun.  For this range
of companion masses, the range of neutron star masses is 
1.2-1.8\Msun.  Accurate determination of $q$, which may be possible
via measurements of the rotational broadening of the secondary 
or light curve modeling, is critical to improve the mass constraint.  
Finding that the neutron star mass is near the upper end of 
the 1.2-1.8\Msun~mass range would have important implications for 
neutron star equations of state, and, in general, determining the mass 
has important implications for the evolution of neutron star systems.

The source distance we obtain is consistent with previous estimates 
\citep{homan99,tomsick99}, confirming that XTE~J2123--058 is unusually 
far from the Galactic plane and implying a bolometric luminosity 
between $3\times 10^{37}$ and $10^{38}$~erg~s$^{-1}$
for the brightest X-ray bursts detected during the 1998 outburst
\citep{tomsick99}.  Given the distance and longitude ($l = 46.5^{\circ}$) 
of XTE~J2123--058, the systemic radial velocity of $-94.5$~km~s$^{-1}$ 
is significantly different from what would be observed for an
object corotating with the disk of the Galaxy.  This fact alone 
does not allow us to distinguish between a scenario where the 
source was ejected from the Galactic plane and the possibility 
that XTE~J2123--058 has a halo origin.  

\acknowledgements

We wish to thank John O'Meara for his help with observations and data reduction.  
We thank Tom Barlow for assistance with his makee software.  We acknowledge Judith 
Cohen for obtaining a standard star exposure for us and Bob Goodrich for his 
help with ESI.  We thank Gary Puniwai and Charles Sorenson for assistance.  
Data for this work were obtained at the W.M. Keck Observatory, which is operated 
as a scientific partnership among the University of California, Caltech and NASA.

% BIBLIOGRAPHY

\end{document}